\documentclass[A4,11pt]{article}
\usepackage{times}
\usepackage{graphicx}
\usepackage{hyperref}

\newcommand{\mtwonui}{\mbox{$m^2(\nu_{\mathrm i})$}}

\newcommand{\mnui}{\mbox{$m(\nu_{\mathrm i})$}}

\newcommand{\mnue}{\mbox{$m(\nu_{\mathrm e} )$}}

\newcommand{\mtwonue}{\mbox{$m^2(\nu_{\mathrm e} )$}}

\newcommand{\mee}{\mbox{$m_\mathrm{ee}$}}

\newcommand{\ttwo}{\mbox{$\rm T_2$}}

\newcommand{\kr}{\mbox{$\rm ^{83m}Kr$}}
\newcommand{\rhenium}{\mbox{$\rm ^{187}Re$}}
\newcommand{\holmium}{\mbox{$\rm ^{163}Ho$}}

\newcommand{\ev}{\mbox{$\rm eV/c^2$}}

\newcommand{\bdec}{\mbox{$\beta$-decay}}
\newcommand{\bspec}{\mbox{$\beta$-spectrum}}
\newcommand{\belec}{\mbox{$\beta$-electron}}

\newcommand{\ezero}{\mbox{$E_0$}}

\newcommand{\be}{\begin{equation}}
\newcommand{\ee}{\end{equation}}
\newcommand{\bea}{\begin{eqnarray}}
\newcommand{\eea}{\end{eqnarray}}

\newcommand{\etal}{\mbox{\it et~al.}}

\chardef\bslash=`\\ 

\begin{document}
\title{Neutrino Masses}

\author{Christian Weinheimer$^{a}$, Kai Zuber$^{b}$\\[2mm]
{\small $^a$ Institut f\"ur Kernphysik, Westf\"alische Wilhelms-Universit\"at M\"unster,}\\
{\small Wilhelm-Klemm-Str. 9, D-48149 M\"unster, Germany}\\
{\small E-mail: {\sf weinheimer@uni-muenster.de}}\\[2mm]
{\small $^b$ Institut f\"ur Kern- und Teilchenphysik, Technische Universit\"at Dresden,}\\ 
{\small Zellescher Weg 19, D-01069 Dresden, Germany}\\
{\small E-mail: {\sf zuber@physik.tu-dresden.de}}\\
}

\maketitle

\begin{abstract}
The various experiments on neutrino oscillation evidenced  that neutrinos have indeed non-zero masses but cannot tell us the absolute neutrino  mass scale. This scale of neutrino masses is very important for understanding the evolution and the structure
formation of the universe as well as for nuclear and particle physics beyond the present Standard Model. 
Complementary to deducing constraints on the sum of all neutrino masses from cosmological observations two different methods
to determine the neutrino mass scale in the laboratory are pursued: the search for neutrinoless double \bdec\ and the 
direct neutrino mass search by investigating single $\beta$-decays or electron captures. The former method is not only sensitive to neutrino masses but also probes the Majorana character of neutrinos and thus lepton number violation with high sensitivity. Currently quite a few experiments with different techniques are being constructed, commissioned or are even running, which aim for a sensitivity on the neutrino mass of  {\cal O}(100)~meV. The principle methods and these experiments will be discussed in this short review.
\end{abstract}

\maketitle

\section{Introduction}

The various experiments with atmospheric, solar, accelerator and reactor neutrinos  provide compelling evidence that neutrino flavor states are non-trivial superpositions of neutrino mass eigenstates and that neutrinos oscillate from one flavor state into another during flight. By these neutrino oscillation experiments the neutrino mixing matrix $U$ containing the mixing angles as well as the differences between the squares of neutrino masses can be determined \cite{fogli12}.

The value of the neutrino masses are very important for astrophysics and cosmology to describe the role of neutrinos in the evolution of the universe. Although neutrinos are very light they may contribute significantly to the mass density of the universe: With 336 neutrinos per cm$^3$ left  over from the big bang they are about a billion times more abundant than atoms. 
On the other hand the values and the pattern of the neutrino masses are very important for nuclear and particle physics, since 
they are a very sensitive probe for physics beyond the Standard Model of particle physics at large scales: Since neutrinos are neutral there is the possibility that neutrinos are their own antiparticles and, additionally, so-called Majorana mass terms originating from large scales could play the dominant role in describing neutrino masses \cite{numass_theory}.

Clearly, neutrino oscillation experiments prove that neutrinos have non-zero masses, but they  -- being a kind of {\it interference experiment} --
cannot determine absolute masses. 
Therefore, we need other ways to determine the absolute value of the neutrino masses. Three methods are sensitive to the values of the neutrino mass eigenstates and their mixing angles in different ways:

\subsection{Neutrino mass from cosmology}
The relic neutrinos would have smeared out fluctuation on small scales, depending on their mass. 
  By analysing the power spectrum of the universe limits on the sum of the three neutrino mass states, e.g.
   $\sum \mnui < 0.5$~eV \cite{hannestad12}, 
   have been obtained which are to some extent model and analysis dependent.

\subsection{Neutrino mass from neutrinoless double \bdec\ ($0 \nu \beta \beta $)}

\begin{figure}[ht!]
\centerline{\includegraphics[width=0.9\textwidth]{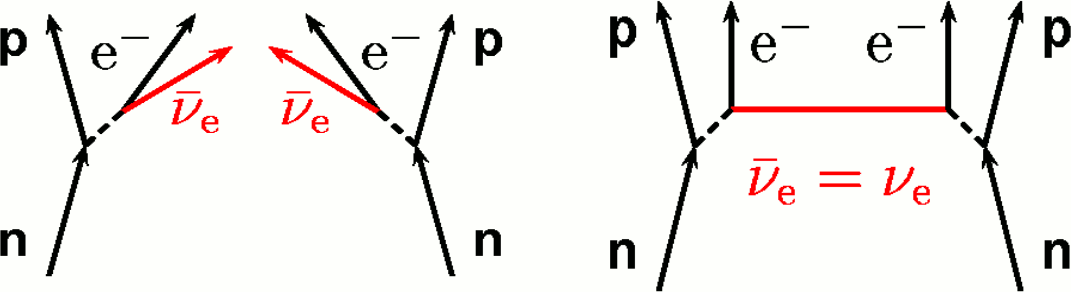}}    
\caption{Normal double \bdec\ with the emission of two antineutrinos (left) and neutrinoless double \bdec\ (right). The diagrams are shown for the case
of a $\beta^-\beta^-$ decay.
\label{fig:dbd_diagram}} 
\end{figure}

  Some even-even nuclei can only decay via double  \bdec\ into a nucleus with higher binding energy. This 2$^\mathrm{nd}$ order week process
  has been proposed more than 70 years ago \cite{goeppert_mayer} and has been experimentally confirmed for around a dozen of nuclei since more than 20 years (see figure \ref{fig:dbd_diagram} left). If -- in the case of a $\beta^-\beta^-$-decay -- the electron antineutrino going out at one vertex is absorbed at the other vertex as neutrino (see figure \ref{fig:dbd_diagram} right) the double \bdec\ will be neutrinoless. It would violate lepton number conservation by two units. Therefore, neutrinoless double {\bdec}   is forbidden in the Standard Model of particle physics. It could exist only, if the neutrino is its own antiparticle (``Majorana-neutrino'' in  contrast to ``Dirac-neutrino''). Secondly, the left-handedness of neutrinos and the right-handedness of antineutrino in charge current weak interactions provide a $2^\mathrm{nd}$ obstacle for neutrinoless double \bdec .  A finite neutrino mass is the most natural explanation   to produce in the chirality-selective weak interaction a neutrino with a small component of opposite handedness on 
  which this neutrino exchange subsists. Then the decay rate will scale with the absolute square of the so called effective neutrino mass,
  which takes into account the neutrino mixing matrix $U$:
  \be\
   \label{eq:mee}
    \Gamma_{0\nu\beta\beta} \propto \left| \sum U^2_{\rm ei} \mnui \right|^2 := \mee^2
  \ee
  In case of neutrinoless double \bdec\ the neutrino mixing matrix $U$ also contains 2 so-called Majorana-phases in addition to the normal CP-violating
  phase $\delta$. The latter is important for neutrino oscillation whereas the former 
  do not influence neutrino oscillation but \mee . A significant additional uncertainty
  entering the relation of \mee\ and the decay rate comes from the uncertainties of the nuclear matrix elements 
  of the neutrinoless double \bdec\ \cite{simkovic09}. 

In case of $\beta^+\beta^+$ decays there are two alternative processes including one or two electron capture (EC) processes: $\beta^+\mathrm{EC}$ and ECEC. However, the modes involving positrons are phase-space suppressed and only six possible $\beta^+\beta^+$ emitters are known. 
  Since in case of neutrinoless double \bdec\ the inner neutrino propagator is not observable the exchange could subsists on a completely different particle allowing this lepton number violating process, e.g. a particle from theories beyond the Standard Model, which leads to a very interesting interplay with new LHC data \cite{rodejohann2011}, because at the TeV scale their contribution to double beta decay can have  a similar amplitude 
 then the light neutrino exchange. Among others there are heavy Majorana neutrinos, right-hand W-bosons and double charged higgs boson,
 which are getting constrained by measurements of ATLAS and CMS \cite{atlas12a,atlas12b,cms12a,cms12b}. But there is a general theorem, that there will be always a Majorana neutrino mass term in case neutrinoless double \bdec\ will observed \cite{schechter_valle}. Diagrams like the one shown in fig. \ref{fig:dbd_diagram} right can also occur for other out-going leptons in theories beyond the Standard Model \cite{zuber_lobster}.

There are many recent reviews on neutrinoless double \bdec\ and neutrinoless double \bdec\ searches, e.g. \cite{rodejohann2011,schwingenheuer2013}.

\subsection{Neutrino mass from direct neutrino mass determination}
  The direct neutrino mass determination is based purely on kinematics or energy and momentum conservation without further assumptions. 
    In principle there are two methods: time-of-flight measurements and precision investigations of weak decays.
  The former requires very long baselines and therefore
  very strong sources, which only cataclysmic astrophysical events like a core-collapse supernova could
  provide.  From the supernova SN1987a in the Large Magellanic Cloud upper limits of 5.7~\ev\  (95~\%~C.L.) \cite{loredo02} or 
  of 5.8~\ev\ (95~\%~C.L.) \cite{pagliaroli10} on the neutrino mass have been deduced, which depend somewhat on the underlying supernova model.  
  Unfortunately nearby supernova explosions are too rare
  and seem to be not well enough understood to compete with the laboratory direct neutrino mass experiments.

  Therefore, the investigation of the kinematics of weak decays and more explicitly the
  investigation of the endpoint region of a \bdec\ spectrum (or an electron capture) is 
  still the most sensitive model-independent and direct method 
  to determine the neutrino mass. 
  Here the neutrino is not observed but the charged decay products are precisely measured. Using energy and momentum conservation 
  the neutrino mass can be obtained. In the case of the investigation of a \bspec\ usually the ``average electron neutrino mass squared''
  \mtwonue\ is determined \cite{otten08}:
  \be\ \label{eq:define_mnue}
    \mtwonue := \sum |U^2_{\rm ei}| \mtwonui
  \ee\ 
   This incoherent sum is not sensitive to phases of the neutrino mixing matrix in contrast to neutrinoless double \bdec .
   
\begin{figure}[t!]
\centerline{\includegraphics[width=0.8\textwidth]{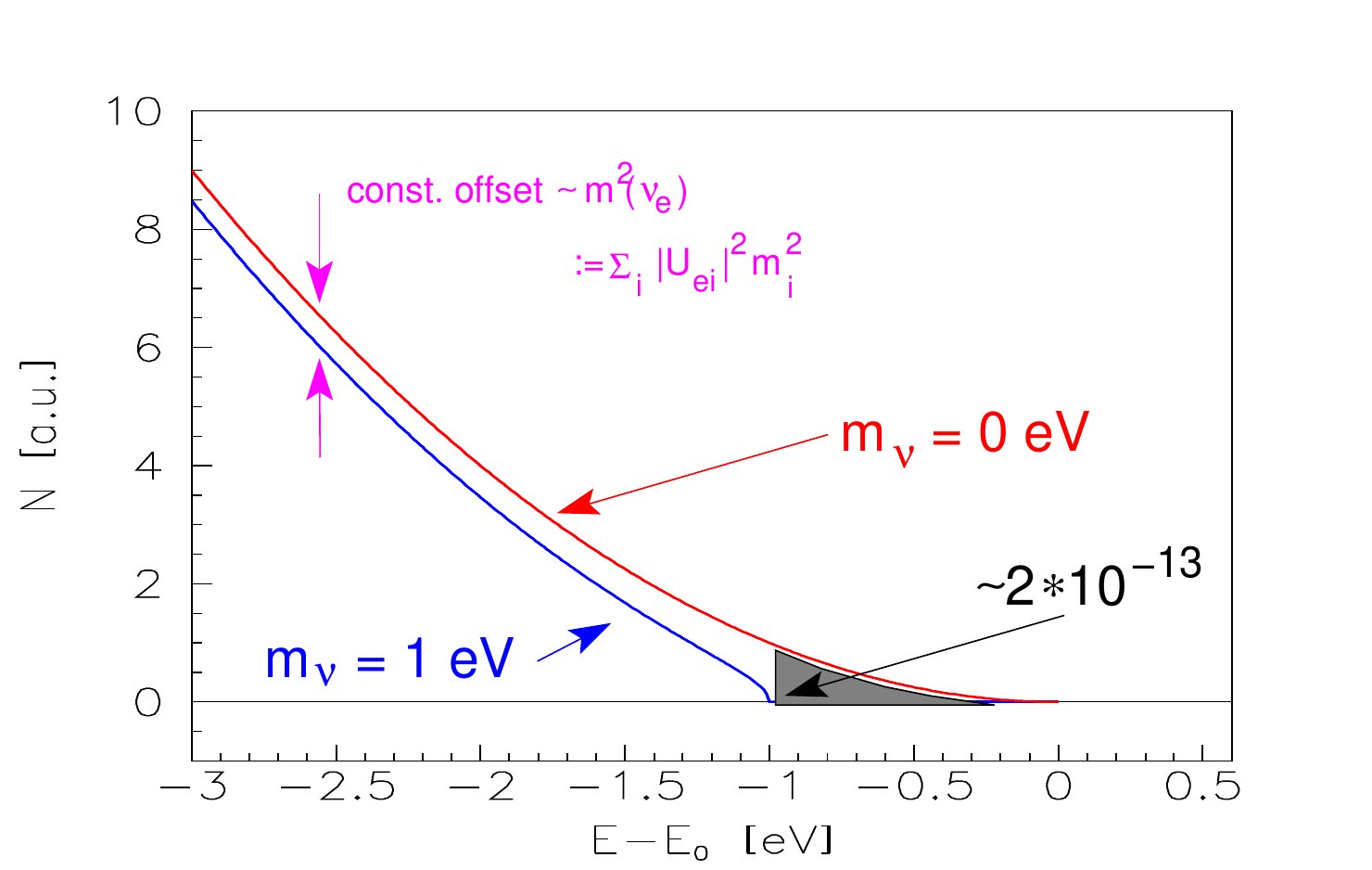}}
\caption{Expanded \bspec\ of an allowed or super-allowed \bdec\ around its endpoint \ezero\
   for $\mnue = 0$ (red line) and for an arbitrarily chosen neutrino mass
  of 1~eV (blue line).
  In the case of tritium \bdec , the gray-shaded area corresponds to a fraction of $2 \cdot 10^{-13}$ of all
  tritium \bdec s.
\label{fig:beta_spec}}
\end{figure}

  In \bdec\, e.g. $(A,Z) \rightarrow (A,Z') + e^- + \bar \nu_\mathrm{e}$ the outgoing electron is sharing the decay energy with the outgoing electron antineutrino. Therefore the shape of the \bspec\ near its endpoint \ezero , i.e. the maximum energy of the electron in case of zero neutrino mass, is sensitive 
to the neutrino mass as shown in figure \ref{fig:beta_spec}. A recent review about this topic is reference \cite{drexlin13}.

\subsection{Comparison of the different neutrino mass methods}

\begin{figure}[tb]
\centerline{\includegraphics[width=0.49\textwidth]{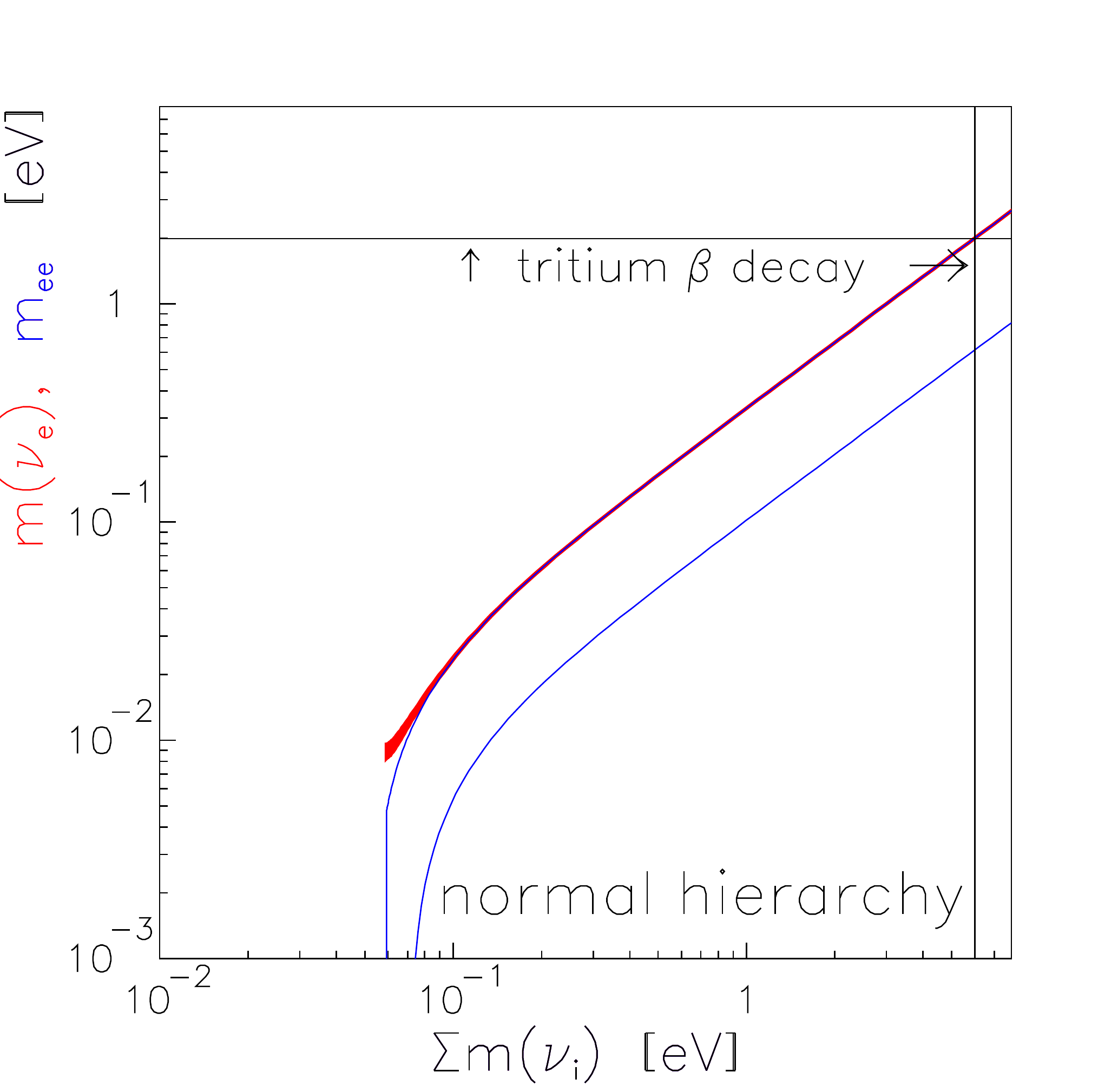}    
\includegraphics[width=0.49\textwidth]{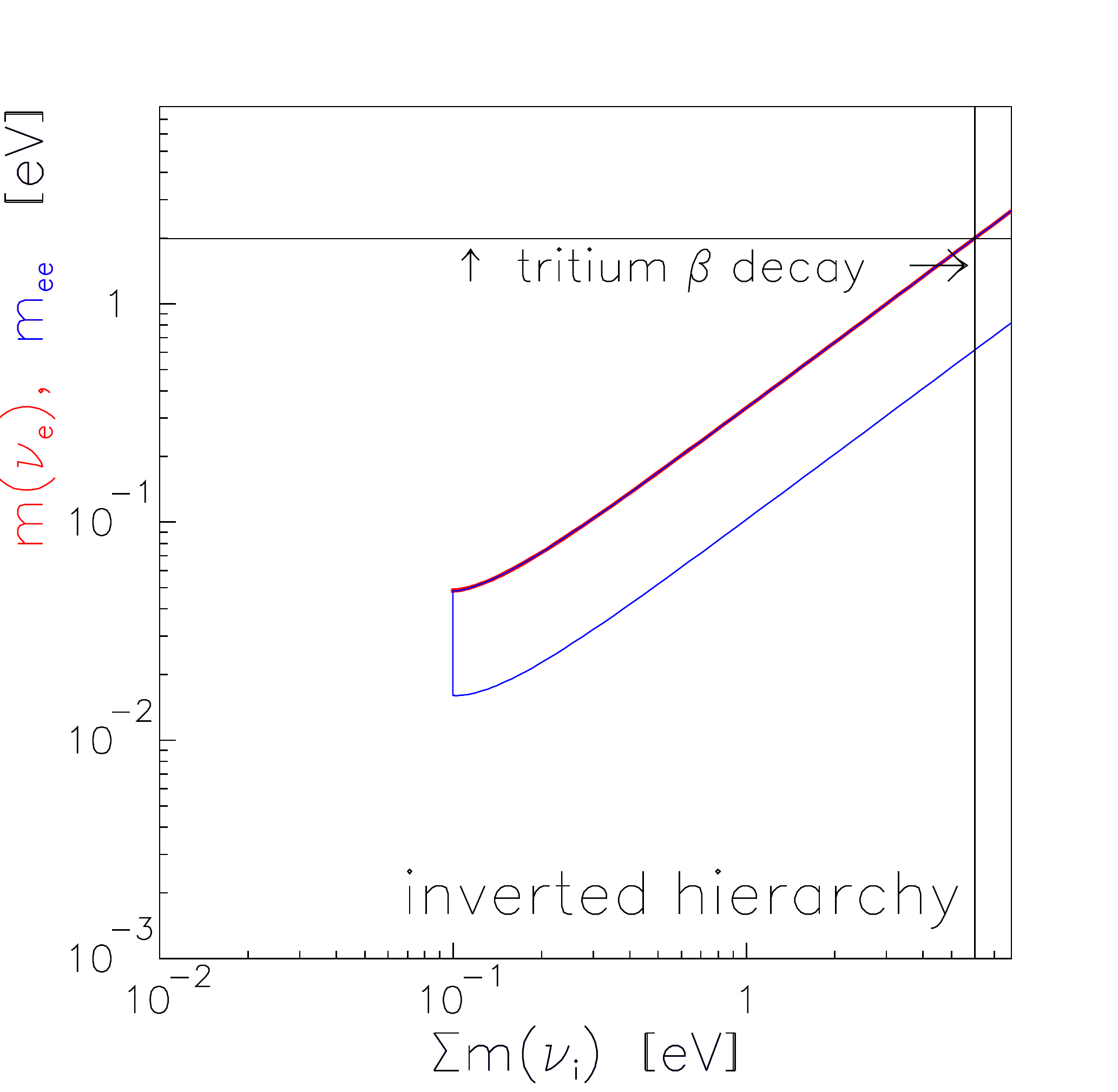}}
\caption{Observables of neutrinoless double \bdec\ \mee\ (open blue band) and of direct neutrino mass determination by single \bdec\ \mnue\ 
(red) versus the cosmologically relevant sum of neutrino mass eigenvalues 
$\sum \mnui$ for the case of normal hierarchy (left) and of inverted hierarchy (right). 
The width of the bands/areas is caused by the experimental uncertainties ($2 \sigma$)
of the neutrino mixing angles \cite{fogli12}
and in the case of \mee\ also by the completely unknown Majorana-CP-phases. 
Uncertainties of the nuclear matrix elements, which enter the determination of \mee\ from the  measured values of half-lives or of half-live limits of neutrinoless double \bdec , are not considered.
\label{fig:comparison_methods}} 
\end{figure}

Figure \ref{fig:comparison_methods} demonstrates that the different methods are complementary to each other and 
compares them. It shows, that the cosmological relevant neutrino mass scale  $\sum \mnui$ has a nearly 
full correlation to \mnue\ determined by direct neutrino mass experiments. The observable of 
neutrinoless double \bdec , the effective neutrino mass \mee ,
does not allow a very precise neutrino mass determination, {\it e.g.} to determine $\sum \mnui$ , due to the
unknown CP and Majorana phases and the uncertainties of the nuclear matrix elements \cite{simkovic09}. 
In the case of normal hierarchy and small neutrino masses the effective neutrino mass \mee\ even can vanish (see figure \ref{fig:comparison_methods} left), which is not possible in the case of inverted hierarchy (see \ref{fig:comparison_methods} right).
On the other hand the combination
of all three methods gives an experimental handle on the Majorana phases. 
As already mentioned in addition the exchange of SUSY particles may be the dominant process of neutrinoless double \bdec , which would spoil the
whole information on the neutrino mass. Nevertheless, the search for the neutrinoless
double \bdec\ is the only way to prove the Majorana character of neutrinos and one of the most promising ways to search for lepton number violation. 

This article is structures as follows: Section 2 reports on the various searches for neutrinoless double \bdec . 
In section 3 the neutrino mass determination from tritium and \rhenium\ \bdec\ as well as from \holmium\ electron capture are presented.
The conclusions are given in section 4.

\section{Search for neutrinoless double $\bf \beta$ decay}
There are  35 double \bdec\ isotopes with the emission of two electrons, the strong dependence of the phase 
space with the Q-value only makes 11 of them (Q-value larger than 2 MeV) good candidates. 
For most of them the normal double \bdec\ with neutrino emission has been observed. 
For neutrinoless double \bdec\ there is only one claim for evidence at $\mee \approx 0.3 ~\ev$ by part of the Heidelberg-Moscow 
collaboration \cite{klapdor04,klapdor06}, all other experiments so far set upper limits. A couple of experiments with sensitivity {\cal O}(100)~meV
are being set up to check this claim or started data taking recently. 
Common to all these experiments is the use of ultrapure materials with very little radioactivity 
embedded in a passive and an active shield placed in 
an underground laboratory. Most of them are using isotopical enriched material as well.

\begin{table}[h]
\caption{The table shows the eleven candidate isotopes with a Q-value larger than
2 MeV. Given are the natural abundances and Q-values as determined from precise
Penning trap measurements. 
The last column shows the experiments addressing the measurement of the corresponding 
isotope. For some experiments only the ''default" isotope is mentioned as they
have the option of exploring several ones. Several additional research and development projects
are ongoing.}
\label{tab:isotopes} 
\begin{center}
\begin{tabular}{|c|c|c|c|}
\hline
Isotope & nat. abund. & Q-value  & Experiment  \\ 
   &    (\%)    & (keV)   & \\  \hline 
  $^{48}$Ca &  0.187& 4262 $\pm$ 0.84 & {\small CANDLES}\\
  $^{76}$Ge & 7.8 & 2039.006 $\pm$ 0.050 &  GERDA, MAJORANA \\ 
  $^{82}$Se & 9.2 & 2997.9 $\pm$ 0.3 &  SuperNEMO, LUCIFER\\
  $^{96}$Zr & 2.8 & 3347.7 $\pm$ 2.2  & - \\
  $^{100}$Mo & 9.6 & 3034.40 $\pm$ 0.17  & AMoRE, LUMINEU, MOON \\
  $^{110}$Pd & 11.8 & 2017.85 $\pm$ 0.64  & -\\ 
  $^{116}$Cd & 7.5 & 2813.50 $\pm$ 0.13 &   COBRA, CdWO$_4$\\
  $^{124}$Sn  & 5.64 & 2292.64 $\pm$ 0.39 &  - \\
  $^{130}$Te & 34.5 & 2527.518 $\pm$ 0.013 &  CUORE \\
  $^{136}$Xe & 8.9 & 2457.83 $\pm$  0.37 & EXO, KamLAND-Zen, NEXT \\
  $^{150}$Nd & 5.6 & 3371.38 $\pm$ 0.20  & SNO+, MCT\\
  \hline
\end{tabular}
\end{center}
\end{table}

The most important signature of neutrinoless double \bdec\ is, that the sum of the energy of both decay electrons 
(in case of double $\beta^-$ decay,  positrons for double $\beta^+$ decay) is equal to the $Q$-value of the nuclear transition. 
The current proposed or running double beta search experiments
are given in Table.~\ref{tab:isotopes}.

Neutrinoless double \bdec\ is also sensitive to different scenarios with sterile neutrinos 
\cite{rodejohann_sterile_nus}. The sum in equation (\ref{eq:mee}) 
will then run over more than 4 neutrino mass states and the corresponding mixing matrix elements. 
The experimental approaches can be classified into two methods (see Figure \ref{fig:dbd_methods}) 
\cite{giuliani10}:

\begin{figure}[t]
\centerline{\includegraphics[width=0.9\textwidth]{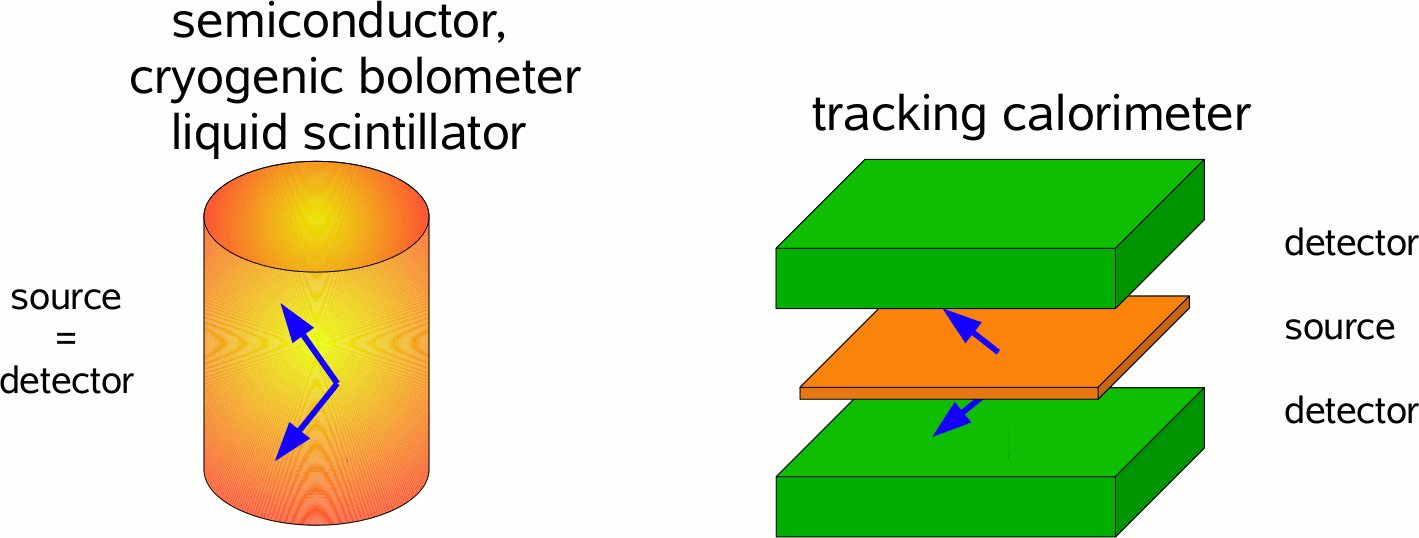}}
\caption{Two different experimental configurations in search for the neutrinoless double \bdec .
\label{fig:dbd_methods}} 
\end{figure}


\subsection{``Source$=$detector'' configuration}
In the ``source$ = $detector'' configuration the double \bdec\ nuclei are part of the detector, 
  which measures the sum of the energy of both \belec s. 
  The experimental implementation of these calorimeters are semiconductors 
   ({\it e.g.} isotopes: $^{76}$Ge, $^{116}$Cd, experiments: GERDA, MAJORANA, COBRA),
  cryo-bolometers ({\it e.g.} isotope: $^{130}$Te, $^{82}$Se, experiments: 
  CUORE, LUCIFER) and liquid scintillators ({\it e.g.} isotope:
  $^{48}$Ca, $^{136}$Xe, $^{150}$Nd, experiments: EXO-200, SNO+, NEXT, KamLAND-Zen, CANDLES).
  In general, this method allows more easily to install a large target mass.

  Currently the most sensitive limits come from the EXO-200 and KamLAND-Zen experiment using
  a large amount of enriched $^{136}$Xe. 
  EXO-200 is a liquid Xenon TPC with a fiducial target mass of 80 kg installed at the WIPP in New Mexico, USA. Coincident drifted charge and scintillation light read-out allows to improve the energy resolution and to reduce the background. EX0-200 gave a half-life limit
  on neutrinoless double \bdec\ \cite{auger2012} of  
  \begin{equation}
  t_{1/2}(^{136}{\rm Xe})  >  1.6 \cdot 10^{25}~{\rm y} \\
   \end{equation}
  The KamLAND-Zen-experiment uses the KamLAND-detector, which was built for long baseline reactor neutrino oscillation measurements, in which a nylon-based inner balloon of 3~m diameter was inserted. 
This balloon is filled with 13~t of Xenon-loaded liquid scintillator.
The scintillation light coming from decays in this balloon is detected by the photomultipliers surrounding 
the KamLAND-detector. For the neutrinoless double \bdec\ search a fidcuial volume with 2.70 diameter containing 179 kg of $^{136}$Xe was used yielding  a half-life limit \cite{gando2013} of
  \begin{equation}
  t_{1/2}(^{136}{\rm Xe})  > 1.9 \cdot 10^{25}~{\rm y}
  \end{equation}
Both the EXO-200 and the KamLAND-Zen results exclude the claimed evidence of part of the Heidelberg-Moscow collaboration for a large part of matrix element calculations.
  
\begin{figure}
\centerline{\includegraphics[angle=0,width=0.9\textwidth]{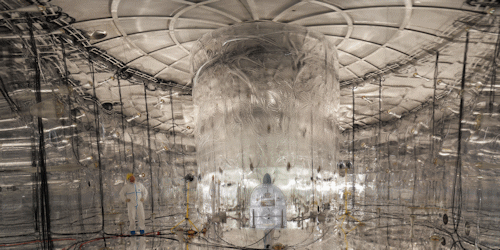}}
\caption{View of the GERDA LAr cryostat within the water shielding which is instrumented as muon veto
(Courtesy of the GERDA collaboration).}
\label{fig:gerda_setup}
\end{figure}

 The GERDA experiment \cite{ackermann2013}
at the Gran Sasso underground laboratory is being proceeded in two phases  with the option of a third phase together with the MAJORANA experiment \cite{majorana10}. GERDA uses enriched Germanium\footnote{The
  enrichment of the double \bdec\ isotope $^{76}$Ge is about 86~\%. The total mass of the phase 1 detectors amounts to 18~kg.}
  embedded in a shielding cryostat filled with liquid argon, which itself sits in a water veto tank (see figure \ref{fig:gerda_setup}). This new shielding technique allowed to improve the background rate compared to the Heidelberg Moscow experiment
  by an order of magnitude. For a second phase point contact BEGe detectors for optimized pulse shape
  analysis are currently produced aiming for another factor 10 in background reduction. 
 
  The GERDA experiment has started data taking in November 2011 and first new results in form of a new
  2$\nu$  double \bdec\ half-life have been obtained \cite{agostini2013}. 
  The GERDA collaboration just recently unblinded their phase I data with a total accumulation 
of 21.6 kg yr \cite{GERDA_phase1_2013}. 
The number of events agrees well with the background expectation.
The experiment sets an lower limit at 90~\% C.L. of the neutrinoless double $\beta$-decay half-life of 
\begin{equation}
	t_{1/2}(^{76}\mathrm{Ge}) > 2.1 \times 10^{25} \mathrm{yr} 
\end{equation}
by GERDA data alone and of
\begin{equation}
	t_{1/2}(^{76}\mathrm{Ge}) > 3.0 \times 10^{25} \mathrm{yr}
\end{equation} 
by using data from former Germanium experiments in addition.
With its about an order of magnitude lower background compared to previous Germanium experiments
the GERDA experiment clearly disfavors the claim by part of the Heidelberg-Moscow collaboration.

 Recently a revived interest for neutrino less double electron capture \cite{bernabeu83} has been grown due to potential 
 resonance enhancement with an excited state of the daughter nucleus \cite{sujkowski2004,Krivoruchenko:2010ng}. 
 Due to the sharpness of the
 resonance a major action was taken with Penning traps to provide better atomic masses and indeed some 
 systems like $^{152}$Gd seem to fulfill the requirement for resonance enhancement (f.e. \cite{eliseev2011}). There is still a lack of understanding what the signal of neutrino less double EC to the ground state could be.

\subsection{``Source$\neq$detector'' configuration}
In the this configuration the double \bdec\ source is separated from two tracking calorimeters, which determine direction
 and energy of both \belec s separately ({\it e.g.} isotope $^{82}$Se, $^{100}$Mo, experiments: NEMO3 and its much larger 
  successor SuperNEMO, ELEGANT, MOON).

  By this method the most sensitive limit comes from the NEMO3 experiment \cite{simard2012}
  in the Modane underground laboratory LSM. NEMO3 was using thin source foils of a total area of 20~$m^2$. These foils contained 
  7~kg of the double \bdec\
 isotope $^{100}$Mo and 1~kg of the double \bdec\ isotope $^{82}$Se. The foils were surrounded by a tracking chamber 
 in a magnetic field composed of 6400 drift cells working in Geiger mode and calorimeter made out of 1940 plastic scintillators.
 The recent upper limits on neutrinoless double \bdec\ from NEMO3 are \cite{simard2012}:
\begin{eqnarray*}
  t_{1/2}(^{100}{\rm Mo}) & > &1.0 \cdot 10^{24}~{\rm y} \qquad {\rm and} \qquad m_{\rm ee} < 0.31 - 0.96~{\rm eV}\\
  t_{1/2}(^{82}{\rm Se})  & > & 3.2 \cdot 10^{23}~{\rm y} \qquad {\rm and} \qquad m_{\rm ee} < 0.94 - 2.6~{\rm eV}
\end{eqnarray*}

 Although it requires much larger detectors to accumulate similar large target masses as in the ``source$=$detector'' case, there is the advantage,
  that the independent information of both electrons allows to study double \bdec\ processes with 2 neutrinos in detail. In case neutrinoless
  double \bdec\ would be detected, the angular correlation of both electrons will allow to draw some conclusion on the underlying process 
  \footnote{A theorem by Schechter and Valle \cite{schechter_valle} requests the neutrinos to have non-zero Majorana masses, 
  if neutrinoless double \bdec\ is proven to exist, but the dominant process could still be different, {\it e.g.} based on other BSM physics like right-handed weak charged currents, which would show a completely different angular
  distribution of the two electrons with respect to a neutrino mass term.}.

\section{Direct neutrino mass experiments}

The signature of a non-zero neutrino mass is a tiny 
modification of the spectrum of the \belec s near its endpoint (see Figure \ref{fig:beta_spec}),
which has to be measured with very high precision. To maximize this effect,
$\beta$ emitters with low endpoint energy (e.g. $\ezero ( \rhenium ) = 2.47$~keV, $\ezero(^3\mathrm{H})= 18.57$~keV) are favored \cite{weinheimer2013}.

\subsection{``Source$\neq$detector'' configuration: Tritium $\beta$ decay experiments}
 
Tritium is the standard isotope for this kind of study due to 
its low endpoint of 18.6~keV, its rather short half-life of 12.3~y,
its super-allowed shape of the \bspec ,
and its simple electronic structure. 
Tritium \bdec\ experiments using a tritium source and a separated 
electron spectro\-meter have been performed in search for the neutrino mass for more than 60~years \cite{otten08,drexlin13}
yielding  a sensitivity of 2~eV by the experiments at Mainz \cite{kraus05} and Troitsk \cite{aseev11}.
The huge improvement of these experiments in the final sensitivity as well as
in solving the former ``negative \mtwonue `` problem with
respect to previous experiments is mainly caused by the new spectrometers
of MAC-E Filter type and by careful studies of the systematics.

\begin{figure}
\centerline{\includegraphics[angle=0,width=1.0\textwidth]{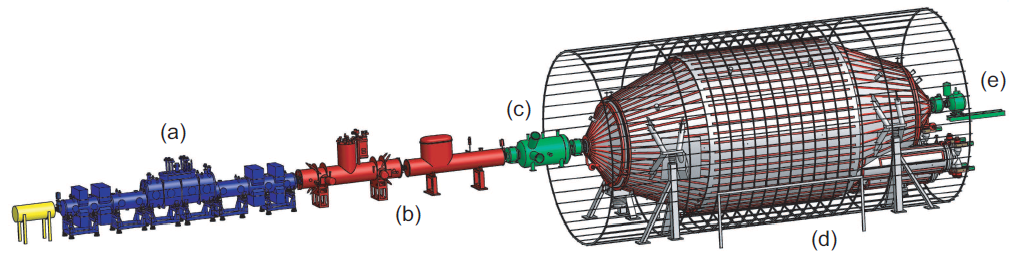}}
\caption{Schematic view of the 70~m long KATRIN experiment consisting of calibration and monitor rear system (yellow), 
windowless gaseous \ttwo -source (a), differential pumping and cryo-trapping section (b), small pre-spectrometer (c) and large main spectrometer (d) and
segmented PIN-diode detector (e). Not shown is the separate monitor spectrometer (Courtesy of the KATRIN collaboration).}\label{fig:katrin_setup}
\end{figure}

To further increase the sensitivity to the neutrino mass down to 200~meV by a direct measurement the KArlsruhe TRItium Neutrino experiment KATRIN \cite{KATRIN_loi,KATRIN_design_report} is currently being set up at the Karlsruhe Institute of Technology KIT.
Since \mtwonue\ is the observable, this requires an improvement by two orders magnitude compared to the previous tritium \bdec\ experiments at Mainz and Troitsk. The KATRIN design is
based on the successful MAC-E-Filter spectrometer technique combined with a very strong windowless gaseous molecular tritium source \cite{KATRIN_design_report}.
Figure \ref{fig:katrin_setup} illustrates the whole 70 m long setup. 

The windowless gaseous molecular tritium source (WGTS) essentially consists of a 10~m long tube of 9~cm diameter kept at 30~K. Molecular tritium gas injected in the middle of this tube is freely streaming to both ends of the beam tube. The tritium gas is pumped back by huge turbo-molecular pumps placed at pump ports intersected with straight 
sections. The \belec s are guided by superconducting solenoids housing the beam tubes. A so-called WGTS demonstrator has been set up to prove that the new concept of the ultra-stable beam-pipe cooling works: gaseous and liquid neon is sent through two tubes welded onto the beam tube. By stabilizing the pressure of this two-phase neon the temperature of the beam tube can be stabilized well below the requirement of $10^{-3}$ \cite{grohmann11}.
The input pressure is chosen to obtain a total column density of $5 \cdot 10^{17}$~molecules/cm$^2$ allowing a near maximum count rate at moderate systematic uncertainties \cite{wgts_monitoring}.
Currently the WGTS demonstrator is being upgraded into the full WGTS.

The electron guiding and tritium retention system consists of a differential and a cryogenic pumping unit. It has been demonstrated that the tritium flow reduction by the differential pumping is about as large as expected by Monte Carlo simulations \cite{lukic11}. Inside the differential pumping sections Fourier transform ion cyclotron resonance Penning traps will be installed to measure the ion flux from the tritium source \cite{ubietodiaz09}. Ions will be ejected from the beam by a transverse electric field. The principle of the cryogenic pumping section based on argon frost at $3 - 4.5$~K has been demonstrated in a test experiment \cite{kazachenko08}. The overall tritium reduction amounts to $10^{-14}$.

A pre-spectrometer will transmit only the interesting high energy part of the $\beta$-spec\-trum close to the endpoint into the main spectrometer \cite{prall12},
 in order to reduce the rate of background-producing 
ionization events therein. The big main spectrometer is  of MAC-E-Filter type as the pre spectrometer. It is essentially an electric retarding spectrometer with a magnetic guiding and collimating field \cite{picard92a}. In order to achieve the strong energy resolution of 1:20,000 the magnetic field in the analyzing plane in the centre of the spectrometer has to be 20,000 times smaller than the maximum magnetic field of 6~T provided by the pinch magnet. Due to conservation of the magnetic flux from the WGTS  to the spectrometer it needs to have a diameter of 10~m in the analyzing plane. To avoid background
by scattering of \belec s inside the spectrometer extreme requirements for the vacuum pressure of $p \approx 10^{-11}$~mbar are necessary \cite{luo07}. 
The \belec s which have enough energy to pass the MAC-E-Filter are counted with a state-of-the-art segmented PIN detector. The spatial information provided by the 148 pixels allow to correct for the residual inhomogeneities of the electric retarding potential and the magnetic fields in the analyzing plane.
Active and passive shields minimize the background rate at the detector. 
 
Of crucial importance
is the stability of the retarding potential. KATRIN is using a twofold way to achieve the necessary redundancy: A custom-made ultra-high precision HV divider 
\cite{thuemmler09} developed together with the PTB Braunschweig and a state-of-the-art 8.5 digit digital voltmeter measure directly the retarding voltage. In addition
the retarding voltage is applied to a third MAC-E-Filter, the so-called monitor spectrometer reusing the former MAC-E-Filter at Mainz.
The line position of ultra-stable electron sources based on the isotope \kr\ \cite{venos10} is continuously compared to the retarding voltage of the main spectrometer.
Both methods reach the required ppm precision.

\begin{figure}
\centerline{\includegraphics[angle=0,width=0.7\textwidth]{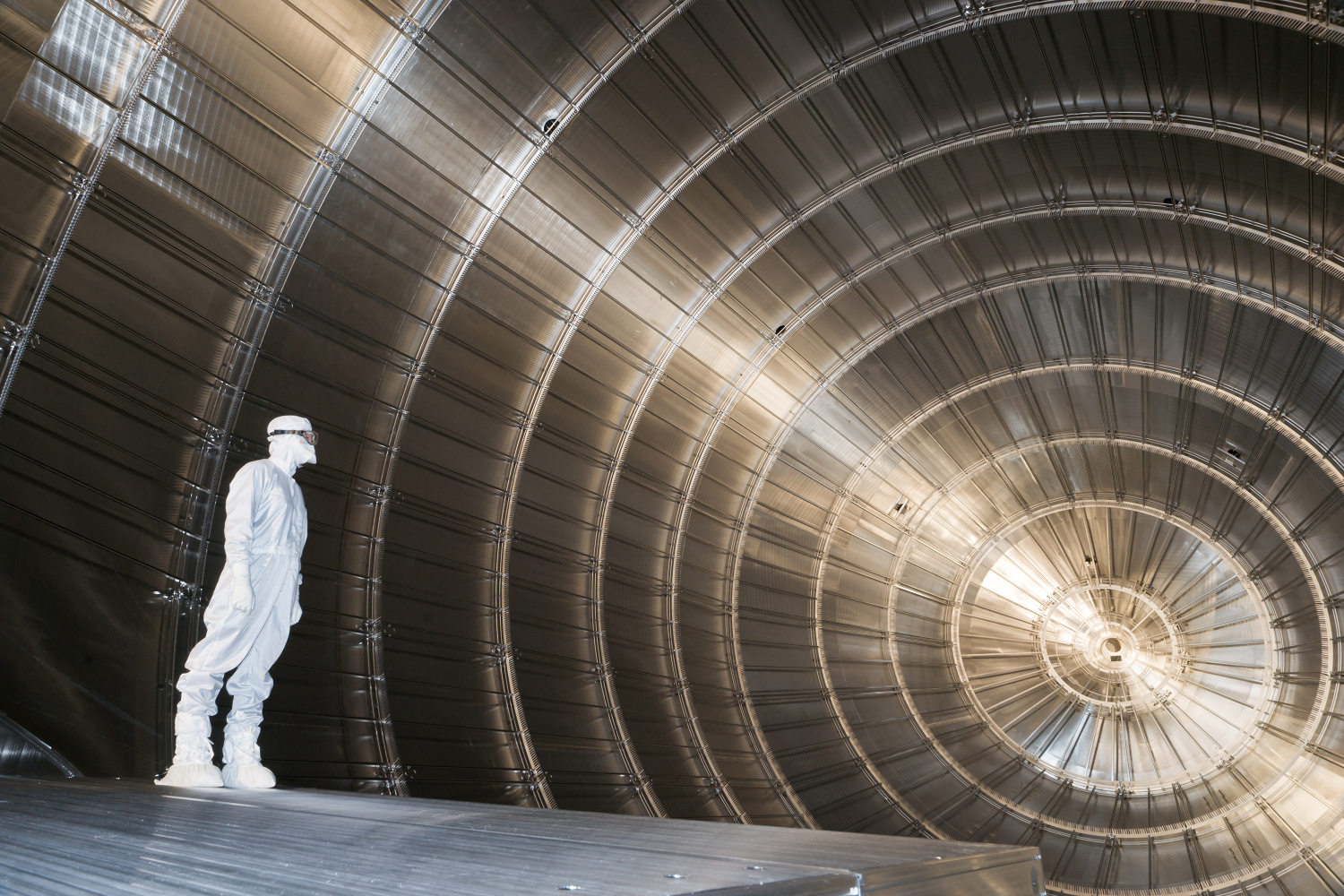}}
\caption{Wire electrode system inside the KATRIN main spectrometer during installation, photo: M. Zacher.}
\label{fig:inner_electrode}
\end{figure}

The sensitivity limit of 200~meV on the neutrino mass for the KATRIN experiment is based on a background rate of $10^{-2}$~cts/s, 
observed under optimal conditions at the experiments at Mainz and Troitsk using similar MAC-E-Filters. To reach this low background rate with the so much larger KATRIN instrument requires new methods.
 At Mainz the main residual background originated from secondary electrons ejected from the walls/electrodes on high potential
by passing cosmic muons or by $\gamma$s from radioactive impurities.
Although there is a very effective magnetic shielding by the conservation of the magnetic flux, small
violations of the axial symmetry or other inhomogeneities allowed a fraction of about $10^{-5}$ of these secondary
electrons to reach the detector and to be counted as background. A new method to reject these secondary electrons from the electrodes has been developed
and successfully tested at the Mainz spectrometer 
\cite{flatt05}: nearly mass-less wires are installed in front of these electrodes, which are put on a more negative electrical potential than the electrode potential by -100~V to -200~V. For KATRIN a double layer wire electrode system consisting of 248 modules with 23440 wires in total has been developed, which should reduce the secondary electron background by a factor 100 \cite{valerius11}. Its installation 
(fig. \ref{fig:inner_electrode}) has been completed in early 2012. 

Other relevant background sources are decays of radioactive atoms in the spectrometer volumes,
e.g. the fast decaying radon isotope $^{219}$Rn from emanation out of the non-evaporable getter pumps \cite{fraenkle11} or small amounts of tritium originating from the WGTS 
\cite{mertens12b}. They create electrons, which might be stored by the magnetic mirror effect and/or by the negative potentials of the two MAC-E-Filters or within the 
non-avoidable Penning trap between the pre and the main spectrometers. For these backgrounds new methods have been developed to avoid storage of electrons or to eject them \cite{beck10,hillen11,mertens12c}. 

Since the KATRIN experiment will investigate only the very upper end of the \bspec , quite a few
systematic uncertainties will
become negligible because of excitation thresholds. Others systematics like the inelastic scattering fraction or the source intensity
will be controlled very precisely by measuring the column density online by an angular-selective electron gun \cite{valerius11,hugenberg10}, by 
keeping the temperature and pressure within the tritium source at the per mille level constant \cite{wgts_monitoring} and by determining the tritium 
fraction of the gas in the source by laser Raman spectroscopy to the sub per mille level \cite{sturm10}. An important consistency check of the correct systematic corrections will be comparison of the endpoint energy \ezero\ fitted from the \bspec\ with a precision value derived from 
ultra-high precision ion cyclotron resonance mass spectroscopy in a multi-Penning trap setup measuring the $^3$He-$^3$H mass difference \cite{blaum2010} with the final goal to use the measured $Q$-value in the neutrino mass fit.
 KATRIN's sensitivity will allow to fully investigate the {\it quasi-degenerate neutrino mass} regime to distinguish between different neutrino mass models as well
as to fully investigate the cosmological relevant neutrino mass range, where neutrino masses
would shape significantly structure formation. In addition, the KATRIN experiment will be sensitive to contributions to sterile neutrinos
\cite{riis11,formaggio11a,Esm12} as suggested by the so-called reactor anomaly.

The commissioning of the KATRIN spectrometer and detector system has started in May 2013. The tritium source as well as the electron transport and tritium elimination section will be put into operation in 2014. First tritium data with the full KATRIN setup are expected for 2015.

There is also R\&D on rather different approaches, like Project-8, which wants to measure
the endpoint spectrum of tritium \bdec\ by detecting the radio emission of coherent cyclotron radiation from a KATRIN-like tritium source \cite{monreal09,formaggio11}. Its main idea is that the cyclotron frequency 
$\omega = (eB)(\gamma m_e)$ scales inversely with $\gamma$,  only the radiated power but not the frequency depends on the angle of the emitted \belec\ w.r.t. the magnetic field. Measuring the \bspec\
by synchrotron radiation has the principle advantage that the radiofrequency photons can leave a tritium
source, which is already opaque for electrons thus allowing much larger source strengths. Currently
the Project-8 collaboration is investigating, whether this very low intensity coherent cyclotron radiation can be detected.

\subsection{``Source$=$detector'' configuration: $^{187}$Re $\beta$-decay and $^{163}$Ho electron capture experiments} 
\subsubsection{$^{187}$Re $\beta$-decay experiments}
Compared to tritium the isotope \rhenium\ has a 7 times lower endpoint energy of 2.47~keV resulting in a 350 times higher relative fraction
of the \bspec\ in the interesting endpoint region. Unfortunately \rhenium\  exhibits a very complicated electronic structure and a 
very long half life of $4.3 \cdot 10^{10}$~y. This disadvantage can be compensated by using it as $\beta$-emitter in cryo-bolometers,
which measure the  entirely
energy released in the absorber, except that of the neutrino. 

A cryo-bolometer is not an integral spectrometer 
like the MAC-E-Filter but measures always the entire \bspec .
Pile-up of two random events may pollute the endpoint region of a \bdec\ on which the neutrino mass is imprinted. 
Therefore cryo-bolometers with mg masses are 
required to suppress pile-up by 4 or more orders of magnitude. Unfortunately 
large arrays of cryo-bolometers are then required to reach the necessary sensitivity to the neutrino mass. 
Another technical challenge is the energy resolution of  the cryo-bolometers. Although cryo-bolometers with an energy resolution of a few eV have been produced
with other absorbers, this resolution has  not yet been achieved with rhenium.

Two groups have started the field of \rhenium\ \bdec\ experiments:
The MANU experiment at Genoa was using one metallic rhenium crystal of 1.6~mg working at a temperature of 100~mK and 
read out by Germanium doped thermistor. The $\beta$ environmental fine structure was observed for the first time
giving rise to a modulation of the shape of the \bspec\ by the interference of the out-going \belec\ wave with the
rhenium crystal \cite{gatti99}. The spectrum near the endpoint allowed to set
an upper limit on the neutrino mass of $\mnue < 26$~eV \cite{gatti01}. 
The MiBeta collaboration at Milano was using 10 crystals of AgReO$_4$ with a mass of about 0.25~mg each \cite{sisti04}. The energy resolution
of a single bolometer was about 30~eV. One year of data taking resulted in an upper limit 
of $\mnue < 15$~eV \cite{sisti04}.

Both groups are now working together with additional groups in the MARE project \cite{mare06}
to further the development of sensitive mi\-cro-ca\-lori\-me\-ters investigating the \rhenium\ \bdec .
MARE consists of two phases \cite{nucciotti08}: MARE-1 aims to investigate alternative micro-calorimeter concepts to improve the energy resolution, to shorten the rise time of the signals and to develop possibly a multi-plexing read-out. A summary of the sensitivity reach dependent on these detector properties can be found in \cite{nucciotti10}.
Among these possible technologies for MARE are transition edge and neutron-doped thermistors for the
temperature read-out, but also new technologies based on magnetic micro-calorimeters \cite{ranitzsch12}.
These new dectectors are being tested in medium-size arrays with up to 300 cryo-bolometers enabling MARE-1
 to reach a sensitivity to the neutrino mass of a few eV. 
After selection of the most successful technique a full scale experiment with sub-eV sensitivity to the neutrino mass will then be set up in MARE phase 2
comprising about 50000 detectors.

\subsubsection{$^{163}$Ho electron capture experiments}
MARE is not only aiming at the \rhenium\ \bdec\ but also wants to investigate the electron capture of \holmium , triggered by the persisting difficulties with superconducting metallic rhenium absorbers coupled to the sensors~\cite{ferri12}. The isotope \holmium\ could be implanted into well-suited cryo-bolometers. 
The very upper end of the electromagnetic de-excitation spectrum of the \holmium\ daughter $^{163}$Dy looks similar to the endpoint spectrum of a \bdec\ and is sensitive to the neutrino mass \cite{derujula}. Additionally, 
the ECHO collaboration has been set up to investigate the direct neutrino mass search with \holmium\ implanted in magnetic
micro-calorimeters \cite{gastaldo2012}. In
these detectors, the temperature change following an energy
absorption is measured by the change of magnetization of a
paramagnetic sensor material (Au : Er) sitting in an external magnetic field. This change of magnetization is read out by a SQUID.
A first \holmium\ spectrum has been presented \cite{ranitzsch12}.
Again large efforts are been undertaken to develop a multi-plexing read-out technology to allow the run large arrays of these magnetic micro-calorimeters.
\section{Conclusions}

The absolute neutrino mass scale is addressed by three different methods. The analysis of large scale structure data
and the anisotropies of the cosmic microwave background radiation are very sensitive but model dependent. The
search for neutrinoless double \bdec\ requires neutrinos to be their own antiparticles (Majorana neutrinos) and is measuring
a coherent sum over all neutrino masses contributing to the electron neutrino with unknown phases.
Therefore -- even without the contribution of other beyond the Standard Model physics processes -- the value of the neutrino mass cannot be determined very precisely. On the other hand 
the discovery of neutrinoless double \bdec\ would be the detection of lepton number violation, which would
be an extraordinary important discovery.
A few double $\beta$-decay experiments of the second generation like EXO-200,
KamLAND-Zen and GERDA phase I have already delivered exciting 
new data; much more, e.g. from these and other experiments, 
will come in the near future.
Among the various ways to address the absolute neutrino mass scale
the investigation of the shape of \bdec\ spectra around the endpoint is the
only real model-independent method, independently of other beyond the Standard Model physics processes.
 The KATRIN experiment is being setup at Karlsruhe and will start data taking in 2015, whereas
the MARE experiment is commissioning a small array of detectors starting MARE phase 1 and ECHO is developing a new technology of electron capture experiments. 
The latter field of cryogenic calorimeters is also driven by the field of astronomy, where arrays of cryogenic bolometers with \cal{O}(1000) pixels have been established already.
From both laboratory  approaches, the search for neutrinoless double \bdec\ and the direct neutrino mass determination, we expect in the coming years sensitivities on the neutrino mass  of {\cal O}(100)~meV.

\section{Acknowledgments}
The work of the authors is supported by the German Ministery for Education and Research BMBF
and the German Research Society DFG.

\end{document}